\pdfoutput=1
\documentclass{svjour3}                     
\smartqed  
\usepackage{epsfig,amssymb,amsmath}
\usepackage{units}
\usepackage{lineno}
\usepackage{chicaco}
\usepackage{graphicx}

 \journalname{Environmental Earth Sciences}
\begin{document}

\title{Evidence for Anthropogenic Surface Loading as Trigger Mechanism of the 2008 Wenchuan Earthquake}


\author{Christian D. Klose}


\institute{Christian D. Klose \at
           Think Geohazards, New York, USA \\
           \email{christian@cdklose.com}           
}

\date{Received: 22 August 2010 / Accepted: 9 September 2011}

\maketitle

\begin{abstract}
Two and a half years prior to China's M7.9 Wenchuan earthquake of May 2008, at 
least 300 million metric tons of water accumulated with additional seasonal water 
level changes in the Minjiang River Valley at the eastern margin of the Longmen Shan. 
This article shows that static surface loading in the Zipingpu water reservoir 
induced Coulomb failure stresses on the nearby Beichuan thrust fault system at $<$17km depth. 
Triggering stresses exceeded levels of daily lunar and solar tides and  
perturbed a fault area measuring 416$\pm$96km$^2$. These stress perturbations, 
in turn, likely advanced the clock of the mainshock and directed the initial rupture propagation 
upward towards the reservoir on the "Coulomb-like" Beichuan fault with rate-and-state dependent frictional behavior. 
Static triggering perturbations produced up to 60 years (0.6$\%$) of equivalent tectonic loading, 
and show strong correlations to the coseismic slip. Moreover, correlations between 
clock advancement and coseismic slip, observed during the mainshock beneath the 
reservoir, are strongest for a longer seismic cycle (10kyr) of M$>$7 earthquakes. 
Finally, the daily event rate of the micro-seismicity (M$\geq0.5$) correlates well 
with the static stress perturbations, indicating destabilization. 
\end{abstract}

\section{Introduction}
\label{intro}
The Wenchuan earthquake of May 12, 2008 occurred in the Sichuan province of the People's 
Republic of China. This M7.9 event ruptured along the border of the Longmen Shan margin of the 
Tibetan plateau in the West and the Sichuan basin in the East. The earthquake's nucleation point 
was at Long=103.364 and Lat=30.986 at about 16 km depth. The epicentral error is 5 km and the 
focal depth estimation error is 10 to 15 km \cite{Huang-etal2008}.

The Beichuan fault system, consisting of listric and NNW dipping reverse faults, broke 
250-300 km parallel along the Longmen Shan thrust belt (Fig.~1) \cite{Burchfiel-etal2008}.
This intra-continental region has been extensively studied both prior to and after the M7.9 
earthquake in 2008, including a) paleo-seismicity studies 
\cite{Burchfiel-etal2008,Densmore-etal2007,Zhang-etal2009,Zhou-etal2007,Burchfiel-etal1995,Chen-etal1994}, 
b) instrumental recordings of the seismicity prior to and after the mainshock (1970-2009) 
\cite{Hu2007,Lei-etal2008}, c) inversion data analyses of teleseismic body waves of the mainshock 
\cite{Wang-etal2008a,Zhang-etal2009}, d) studies on coseismic ground deformations 
\cite{Lin-etal2008,Zhang-etal2009}. Finally, the trigger mechanism of the 2008 M7.9 Wenchuan earthquake has 
been debated since the earthquake's occurrence \cite{Klose2008,Lei-etal2008}. Some studies suggest that 
the mainshock might have been triggered by pore pressure diffusion within the earth's crust resulting 
from a nearby artificial lake, the Zipingpu water reservoir \cite{Lei-etal2008,Ge-etal2009}. 
Other studies reject the hypothesis of triggering due to pore pressure diffusion \cite{Deng-etal2010,Gahalaut2010}.
It can be anticipated that the earthquake cycle was already in its late stage in 2005 and close 
to failure conditions, because the Wenchuan earthquake ruptured in 2008. 
But, did the surface loading affect the earthquake cycle and the initial rupture? 
How many years of equivalent tectonic loading would the artificial loading advance the clock of 
the mainshock and how much of the coseismic slip would it produce?

This study shows that observations and data modeling support the initial argument that 
lithostatic stress changes and the poroelastic response of the earth's crust due to the weight of 
the Zipingpu reservoir on the earth's crust most likely triggered the 2008 M7.9 Wenchuan earthquake \cite{Klose2008}.
The article provides further evidence that a) surface loading due to water mass accumulations 
within the Zipingpu water reservoir in the Minjiang River Valley  
between 2005 and 2008 induced Coulomb failure stress changes in the earth's crust and b) triggering 
stress perturbations beneath the artificial lake likely advanced the clock of the 
Wenchuan earthquake, while affecting the initial rupture propagation. 
Furthermore, the water reservoir impounding generated static triggering stress perturbations in the earth's 
crust beginning in 2005 that biased daily stress alterations due to tidal elongations of the moon and the sun. 
Furthermore, it could have changed the natural earthquake cycle and advanced 
the clock of the 2008 Wenchuan earthquake by several decades.

\section{Data}
\subsection{Seismicity prior to the mainshock}
Instrumentally recorded seismicity with magnitudes up to 5.0 was observed in this region 
before 2004 \cite{Yang-etal2005Waldhauser,Liu2007,Hu2007,Yao-etal2008,Huang-etal2008}.
Paleoseismic studies also show evidence of Quaternary reverse faulting in the Longmen Shan 
region. Radiocarbon analyses ($^{14}$C), for example, indicate that the last major earthquake 
(M$<$7.9) might have occurred between 4 and 10kyr ago \cite{Densmore-etal2007,Zhou-etal2007}.
Moreover, deformation measurements suggest the Longmen Shan is a transition part between both 
a stable continental region (Sichuan Basin) with strain rates $<10^{-10}$yr$^{-1}$ and an 
active continental region (Tibet plateau) with strain rates $>10^{-8}$yr$^{-1}$.

\begin{center}
\begin{minipage}[h]{12cm}
  \framebox{\includegraphics[width=12cm]{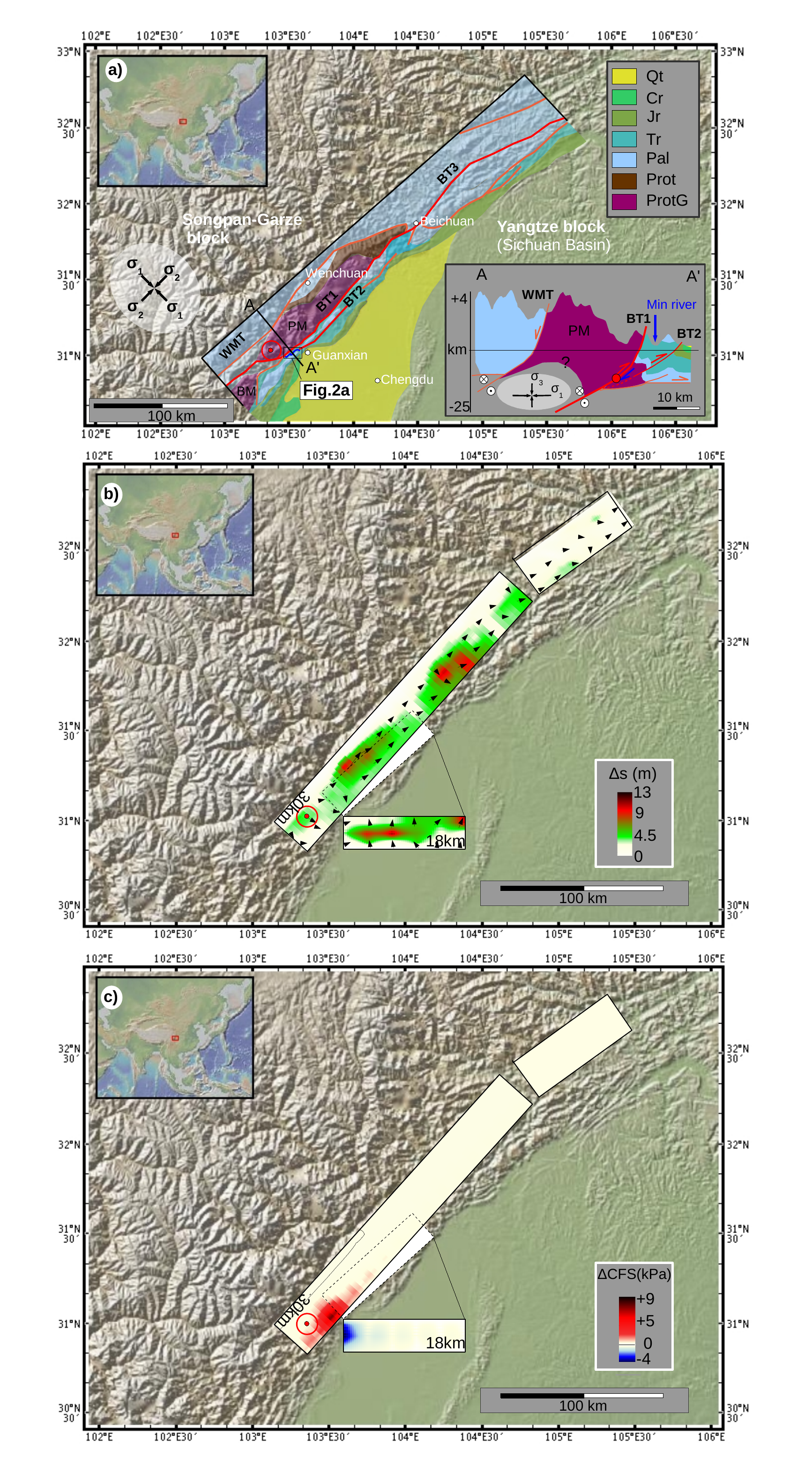}}
{\footnotesize {\bf Fig.~1.} Geology and tectonic situation in the Longmen Shan region (a) 
the Beichuan fault system that ruptured during the May 12 2008 M7.9 earthquake 
with a coseismic slip $s$ \cite{Wang-etal2008a} (b) and likely due to static Coulomb 
failure stress perturbations $\Delta CFS$ (c) as a result of the surface loading in the 
Zipingpu water reservoir (see Fig.~2). The principal lithostatic stress elipsods show the 
orientation of reverse fault contitions. The epicenter and the epicentral error of the mainshock 
are indicated by red circles. The geology is as follows: Qt-Quaternary, 
Cr-Cretaceous, Jr-Jurassic, Tr-Triassic, Pal-Paleozoic, Prot-Proterosoic, 
ProtG-Proterosoic granite, BM-Boashan Massif, PT-Pengguan Massif, 
WMT-Maowen thrust fault (WMT) and BT1/3-Beichuan thrust fault \cite{Burchfiel-etal2008,Jia-etal2006}.}
\end{minipage}
\end{center}

The area where the mainshock nucleated tends to have low horizontal and 
vertical tectonic deformation rates of $\leq$1.0$\pm$1.0 mm yr$^{-1}$, resulting in less seismic 
activities when compared to a) the seismic hazard regions in the Quiangtan block (Southwest) and 
the Kulun Mountains (Northwest) and b) the Mesozoic deformation period 
\cite{Burchfiel-etal2008,Densmore-etal2007,Zhang-etal2009,Zhou-etal2007,Burchfiel-etal1995,Chen-etal1994}.
High shear-wave velocity structures and Bouguer-gravity structures indicate a mechanically 
strong mid crust (10-20 km) and a cratonic-like lithosphere of the Sichuan Basin (Fig.~1a), 
in particular, in the Yangtze block along the SW Longmen Shan at the border to the Songpan-Garze block 
\cite{Yao-etal2008,Burchfiel-etal2008}. Yao et al.~suggested a more unstable lower part of the crust 
($>$20km) along the Longmen Shan which could be explained by low shear-wave velocity structures 
\cite{Yao-etal2008}.

\subsection{Seismicity associated with the Zipingpu reservoir impoundment}
Between October 2005 and May 2008 shallow seismicity patterns ($<$10km) were observed on the 
Beichuan fault system along a $\leq$20 km long part of the lower Minjiang River Valley parallel to the 
extension of the Zipingpu water reservoir \cite{Lei-etal2008,Klose2009}. 
Such shallow seismicity could be indicative for earthquakes in a stable continental crust \cite{KloseSeeber2007}.
However, Lei et al., reported that the seismicity change might be associated with a continuous 
water accumulation by the Zipingpu dam in the Minjiang River Valley almost three years prior to the rupture of 
the mainshock \cite{Lei-etal2008,Klose2008}. At the northern end of the valley, the Minjiang River flows 
into the Chengdu plain of the Sichuan basin. The artificial water reservoir extended sub-parallel 
to the Beichuan fault segment BT1 about 1.5 km westward and BT2 less than 1.0 km eastward 
(Figs.~2 and 1). After continuous impounding, the water level peaked twice in October 2006 and 
October 2007 at water reservoir capacity of 1.10 10$^{9}$m$^3$ (upper water level) which is 
equivalent to a mass of 1.10 billion metric tonnes (Gt). Between October 2007 and May 12 2008 
the reservoir was, again, seasonally drained to a remaining volume of 0.32 10$^{9}$m$^3$ 
(lower water level). Lei et al.~first mentioned that the seismic events of magnitudes ranging 
between 0.5 and 3.9 might have illuminated a destabilization process on BT during the loading 
period of the water reservoir. Prior to the surface loading, the correlation coefficient 
between the water level and the daily event rate (383 days) is -0.18 (Fig.~3). 
Since May 2005 - the start of the flooding season - a positive correlation 
exists between the water level (mass of water) and the micro-seismicity beneath the water 
reservoir. With a coefficient of 0.69, the correlation is strongest between 
2005 and 2006 during the first 366 days of loading. Between 2006 and 2007 the 
correlation is weaker with a coefficient of 0.62. Thus, it could be anticipated that 
the earthquake nucleation processes started already in 2006.

\subsection{Coseismic slip of the mainshock}
Numerical inversion data of teleseismic body waves (P-waves) show that the mainshock 
described a complex rupture process on the Beichuan fault segment BT1 (Fig.~1b) \cite{Zhang-etal2009,Wang-etal2008a}.
During the first 15-20 seconds, a reverse fault focal mechanism dominated at and above the 
nucleation point of the mainshock \cite{Zhang-etal2009,Wang-etal2008a}. The rupture propagated 
with a coseismic slip of up to 4 m on BT1 upwards toward the Minjiang River Valley. 
Thus, it propagated directly to the surface loading area of the Zipingpu water reservoir.
After 20 seconds, the rupture propagation process changed to a right-lateral NNE strike-slip 
mechanism NE of the Minjiang River Valley \cite{Zhang-etal2009,Wang-etal2008a}.

\section{Methods}
\subsection{Triggering stress perturbations caused by surface loading}
\label{sec:Love}
An exact first order solution of stress states below the surface loading area 
(lower and upper water reservoir) in the Minjiang River Valley (Fig.~2) 
is according to Boussinesq's classical solution 
\cite{Boussinesq1885,Love1944}. Boussinesq's solution for point loads is based on the assumption 
that the modulus of elasticity is constant within a homogeneous 3-dimensional half-space 
(earth's crust). Moreover, the principle of linear superposition is also assumed to be valid. 
For the given problem, surface areas of the lower and upper water level with arbitrary 
geometry were discretized in $N$ $a_i \times b_i$ m$^2$ elements (e.g., 50$\times$50 m$^2$).
With respect to the surface elements, A.~Love developed a method to 
analytically determine stress states at a depth $z$ beneath any uniform un/load $L_i$ 
with the area $A_i = a_i \times b_i$ \cite{Love1944}. This analytical solution is free of any elastic 
constant and can be defined for each 2-D surface element $i= 1,2,3,\ldots,N$:

\begin{equation}
\sigma_{L,i}(\Delta m_i,z,a_i,b_i) = L_i(\Delta m_i,A_i) \frac{f(z,a_i,b_i)}{2\pi},\; \text{with} 
 \label{eq-Steinbrenner}
\end{equation}
\begin{equation}
 f(z,a_i,b_i) = \left[ \arctan \left( \frac{a_i\ b_i}{z\ R_i} \right) + \frac{a_i\ b_i\ z}{R_i} \left( \frac{1}{a_i^2 + z^2} + \frac{1}{b_i^2 + z^2} \right) \right],
\end{equation} 

and where $\Delta m_i$ is the mass change and $R_i=\sqrt(a_i^2+b_i^2+z^2)$. $f(z,a_i,b_i)$ describes the fraction of $L_i$ in the vicinity 
beneath $A_i$ and ranges between 0 and 1. The un/loads $L_i$ induce a positive/negative vertical 
stress alteration $\Delta \sigma_3 = \Delta \sigma_{L}$ superimposed over all $\sigma_{L,i}$, 
whereas $\sigma <$ 0 is compression. $L$ also changes the horizontal stress components, due 
to the elastic response of the crust (Hook's Law): $\Delta \sigma_3 = \Delta \sigma_{L}$ and 
$\unitfrac{\nu}{1 - \nu}\, \Delta \sigma_L \leq \Delta \sigma_{1,2} < \Delta \sigma_L$ 
\cite{McGarr1988}, where $\nu$ is the Poisson's ratio. It is likely that 
$\Delta \sigma_{1,2} \geq \unitfrac{\nu}{1-\nu}\, \Delta \sigma_L$, because horizontal principal 
strains $\epsilon_{1,2}$ are very small and can be assumed to be $\epsilon_{1,2} = 0$ in 
11-19 km depth, where faults are generally locked in the Longmen Shan region \cite{Wang-etal2008b}. 

Table \ref{Modelquantities} shows the 3-dimensional model quantities that were taken into account to determine 
the lithostatic stress perturbations in the earth's crust. Stress states were determined 
on 7$\times$4 km$^2$ large fault elements of the double-listric Beichuan fault system 
(BT1,2, and 3) near the artificial reservoir in the Minjiang River Valley. Thus, 
$\sigma_{L,i}$ and the resulting static Coulomb failure stress $\Delta CFS$:

\begin{equation}
\Delta CFS = \Delta \tau_f + \tan\phi\,(\Delta \sigma'_n)
\label{eq:CFS}
\end{equation}

were calculated on each fault element with respect to the lower and upper water level in the 
Minjiang River Valley (Fig.~2), whereby $\sigma'_n$ is the effective normal stress. 
$\Delta CFS$-values were used for further estimations of the clock advancements 
$\Delta t$ with respect to 4 kyr, 7 kyr, and 10 kyr earthquake cycles of the fault system 
(this methodology is described the next section).

Pore-pressure diffusion most likely played a major role for the micro-seismicity (M$<$4) that 
was observed beneath the water reservoir, according to the study of Lei et al.~2008. 
Thus, drained conditions might have existed in shallower depth ($<$10km) in the Paleozoic rocks 
and the permeable Mesozoic rocks. This suggests that the poroelastic 
response due to the lithostatic stress alterations of the surface loading played a minor role 
under these geological conditions. Thus, the Skempton coefficient $B$ was assumed to lie between 
0.5 and 0.7 \cite{Terzaghi1938,Biot1941,RiceCleary1976}.

\begin{equation}
\Delta p = - B \Delta \overline{\sigma},
\label{eq:poroel}
\end{equation}
where $\Delta \overline{\sigma}$ is the change of the mean stress.

In undrained conditions ($>$10km) the pore pressure $p$ increased due to the poroelastic effects 
and destabilized BT1 and BT2 within the low permeable Proterozoic granitic rocks. The poroelastic 
response tends to increase with both depth (confining pressure) and fracture density \cite{LocknerBeeler2003b}.
The Skempton coefficient $B$, which describes the strength of the poroelastic response, was 
assumed to range between 0.7 and 0.8 in 10-23 km depth. It is also assumed that $B$ is 
additionally amplified up to 0.85, due to a higher fracture density \cite{LocknerBeeler2003b} 
near the intersection regions of BT1 and BT2 in 15-17 km and at the South end of BT1 in direction 
to the intersection with the Wenchuan Maowen thrust fault WMT (Fig.~1). Furthermore, the 
Mohr-Coulomb failure criteria indicates that the poroelastic response dominated only on 
steep dipping fault segments ($>$60$^\circ$). Geomechanical parameters, which were used to determine 
triggering stress perturbations, are summarized in Table \ref{Modelquantities}.

\begin{table}[!hb]
	\centering
	\caption{Model quantities for determining the Coulomb failure stress on the Beichuan 
		fault system due to the surface loading in the Minjiang River Valley (Fig.~2).}
\begin{tabular}[t]{lll}
  \hline\hline
  Geomechnaical quantity		  & Mean$\pm$SME	    & Characteristic element on the fault\\\hline\hline
  Poisson's ratio $\nu$:                  & 0.25$\pm$0.05           & Mesozoic rocks ($<$8.5km depths) \\
  Skempton's coefficient $B$ (drained):   & 0.60$\pm$0.10           & Mesozoic rocks ($<$8.5km depths) \\
  Skempton's coefficient $B$ (undrained): & 0.75$\pm$0.05           & Proterozoic rocks ($>$8.5km depths) \\
  Skempton's coefficient $B$ (undrained): & 0.85$\pm$0.25           & faults zones ($>$8.5km depths) \\
  Rock friction angle $\phi$:             & 28$\pm$2$^{\circ}$      & Mesozoic rocks ($<$8.5km depths) \\
  Rock friction angle $\phi$:             & 28$\pm$3$^{\circ}$      & Proterozoic rocks ($>$8.5km depths) \\
  Rock cohesion $c_0$:                    & 10$\pm$10 MPa           & \\
  Rock density $\rho$:                    & 2700$\pm$50 kg m$^{-3}$ & \\
  Young modulus $E$:                      & 75$\pm$25 GPa & \\
  Shear modulus $G$:                      & 30$\pm$10 GPa & \\
  Lithostatic stress regime:                    & $\sigma_1 > \sigma_2 > \sigma_3$ & reverse fault regime \\
  Horizontal lithostatic stresses:              & $\sigma_1$ and $\sigma_2$        &  \\
  Gravitational lithostatic stress $\sigma_3$:  & $\rho\,g\,z$                     & vertical stress \\
  Fault dip angle $\theta$:                     & 20-60$\pm$1$^{\circ}$            & BT1, BT2, and BT3\\\hline\hline
\label{Modelquantities}
\end{tabular}
\end{table}

\subsection{Clock advancement of the mainshock caused by surface loading}
\label{sec:Gomberg}
Paleoseismic analyses show evidence of Quaternary reverse faulting in the Longmen Shan region 
which indicate a M7 (or M8) earthquake recurrence interval of 7$\pm$3 kyr, 
\cite{Densmore-etal2007,Zhou-etal2007} given horizontal and vertical deformation 
rates of $\dot \epsilon$ = 1.0$\pm$1.0 mm yr$^{-1}$ (5$\pm$5\;10$^{-9}$ yr$^{-1}$). The 
rate-and-state dependent friction law \cite{Dieterich1979,Ruina1983}, which was developed 
based on empirical observations and utilized in this study, describes the failure 
time of a preexisting fault (here: Beichuan fault system). Furthermore, such a 
rate-and-state model can estimate the seismic cycle by taking into account a) 
the Mohr-Coulomb failure law, b) the slip velocity of the fault, and c) the history of the 
slip velocity:

\begin{equation}
\tau = \mu,\sigma_n,\; \text{and}\; \mu = f(v, v(t)),\; \text{where} 
\label{eq:rate-state-1}
\end{equation} 
\begin{equation}
\mu = \mu_0 + a\,\ln\left(v(t)/v_0\right) + b\,\ln\left(\xi(t)v_0/d_c\right)
\label{eq:rate-state-2}
\end{equation} 

$a$ and $b$ are dimensionless hyper-parameters and based on empirical 
observations \cite{Blanpied-etal1998}. $\mu_0$ is the initial friction coefficient and 
$v_0$ is the initial slip velocity of the fault. $\xi(t)$ is a time dependent "state" quantity and 
$d_c$ is a critical slip distance \cite{Dieterich1979,Ruina1983}:

\begin{equation}
\frac{d\xi}{dt} = 1 - \xi(t)v(t)/d_c
\label{eq:rate-state-22}
\end{equation}

This model can be used to determine the clock advancement of an earthquake, for example, due to 
static triggering stress perturbations applied during a seismic cycle. 
The static triggering stress perturbation can be approximated by a heavyside function 
(eq.~\ref{eq:rate-state-3}), because the water reservoir impoundment process was abrupt in 2005
(Fig.~2). In detail, the clock advancement $\Delta t$ is the period between the time of failure 
without perturbation $t_f$ and the time of failure when the static triggering load perturbed the 
background \cite{Gomberg-etal1998}.

\begin{table}[h]
	\centering
	\caption{Model quantities for three rate-and state fault regimes describing the potential behavior of the Beichuan fault system.}
\begin{tabular}[t]{lccc}
        \hline\hline
  model parameters & \multicolumn{3}{c}{rate-state model}\\
                   & regime 1     & regime 2     & regime 3\\\hline\hline
  $a$				&0.15&0.005&0.0027\\
  $b$				&0.15&0.15&0.15\\
  $\mu_0$			&0.57&0.57&0.57\\
  $d_c$ \ m			&0.001&0.001&0.001\\\hline\hline
\label{models}
\end{tabular}
\end{table}

\begin{equation}
  \begin{split}
    \Delta t &= t_f - t_p \\
             &= |\Delta \overline{CFS}|/\dot \mu - a/\dot \mu \ln(1 - (L + 1) \exp(-\dot \mu\,t_f/a)),
  \end{split}
\label{eq:rate-state-3}
\end{equation} 

where $|\Delta \overline{CFS}|$ is the Coulomb failure stress of the static perturbation, 
normalized by the associated normal stress $\sigma_n$, and $L$ describes the load function 
respectively:

\begin{equation}
  L = 
  \begin{cases}
  \exp(\dot \mu\,t_o/a)\,(1 - \exp(|\Delta CFS_s|/a)) - 1                    &  \text{if tectonic and static surface load,}\\
  -1									     &  \text{if tectonic load only.}
  \end{cases}
\label{eq:rate-state-4}
\end{equation}

Three rate-state models and three natural earthquake cycles (4kyr, 7kyr, and 10kyr) were 
chosen to estimating the clock advancement of the Wenchuan earthquake by static triggering stress 
perturbations of the artificial water reservoir behind the Zipingpu dam (Fig.~2). The different 
models and cycles will help to better understand the number of years by which the clock of the 
mainshock was advanced. Model parameters are summarized in the following Table \ref{models}. 

\begin{center}
\begin{minipage}[ht]{10cm}
  \framebox{\includegraphics[width=10cm]{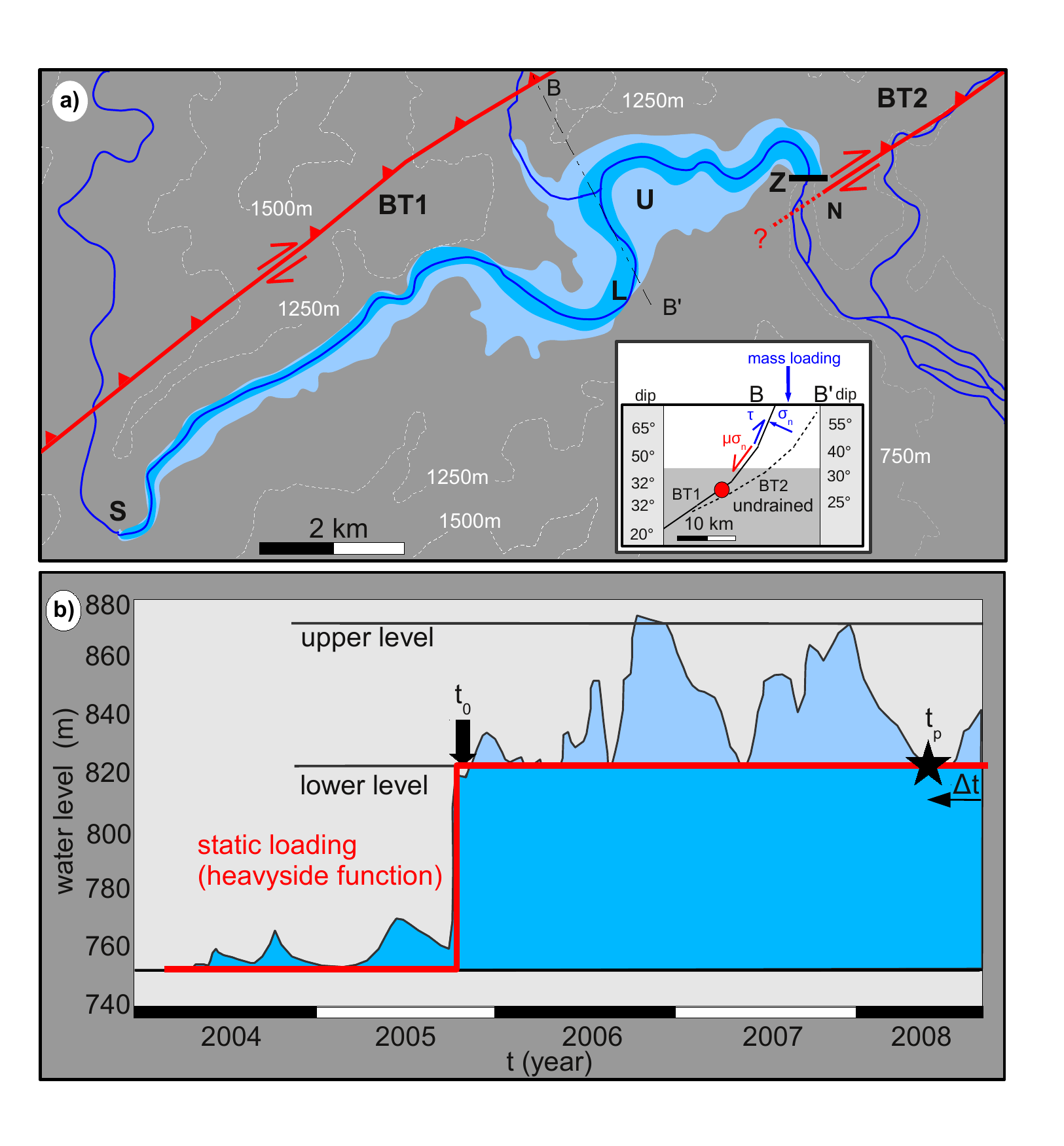}}
{\footnotesize {\bf Fig.~2.} Map view of the Minjiang River Valley and the cross 
section of the double-listric fault model,
including the extension of seasonal lower (L) and upper water level (U) in the valley 
(a). The construction of the Zipingpu dam (Z) was completed in 2004/5 \cite{Lei-etal2008}, 
and followed by the impounding of an artificial water reservoir (b). The loading time is 
indicated by $t_o$ and the mainshock (black star) ruptured at $t_p = t_f - \Delta\,t$, 
while being clock advanced by $\Delta\,t$ year. The water level changed seasonally with a 
wavelength of $t_w=$ 1 year.}
\end{minipage}
\end{center}

\newpage

\section{Results and Discussion}

The period of stress perturbation was about 2.5 years, which accounts for the time between the 
start of the loading $t_o$ in 2005 and the moment of failure $t_p$ in 2008. 
The 3-dimensional stress modeling results show that the static surface loading process 
during the Zipingpu water reservoir impounding altered the lithostatic stresses of the 
reverse fault conditions with a vertical minimum principal stress $\sigma_3$ and a horizontal 
maximum principal stress $\sigma_1$ (Fig.~1a). 
In fact, static stress perturbations brought BT1 closer to failure ($\Delta CFS > 0$) and 
shifted BT2 and the shallow dipping root of BT1 away from failure ($\Delta CFS < 0$) as 
shown in Figure 1c.

First, the water was at lower level (300 Mt) when the mainshock ruptured on May 12, 2008. 
This static load induced shear stresses of $>$1 kPa and normal stresses $>|$-2$|$ kPa (compression, 
if $\sigma<$0) at $<$17 km depth on BT1 beneath the artificial water reservoir 
(see Section \ref{sec:Love}). Given the peak load at upper water level (1.10 Gt), shear 
stresses increased by $>$3 kPa and normal stresses increased by $>|$-6$|$ kPa. Furthermore, 
the shear stress $\Delta \tau_f$ and the effective normal stress $\Delta \sigma'_n$, in turn, 
changed the Coulomb failure stress (see eq.~\ref{eq:CFS}) and brought BT1 closer to failure in 
$<$17 km depth. Again, it should be emphasized that 
$\Delta \sigma'_n$ also changed due to the poroelastic response (destabilization) of BT1  
to the load of the Zipingpu water reservoir (eq.~4). 
BT1 responded, in particular, on steep dipping fault segments ($>$60$^\circ$) within the low permeable 
Proterozoic granitic rocks, as shown in Fig.~3. On the other hand, the influence of the pore pressure diffusion 
is infinitesimal small under undrained conditions in a compressive tectonic regime. 
This fact has been previously discussed \cite{Klose2008,Klose2009,Deng-etal2010,Gahalaut2010}.

\begin{center}
\begin{minipage}[t]{10cm}
  \framebox{\includegraphics[width=10cm]{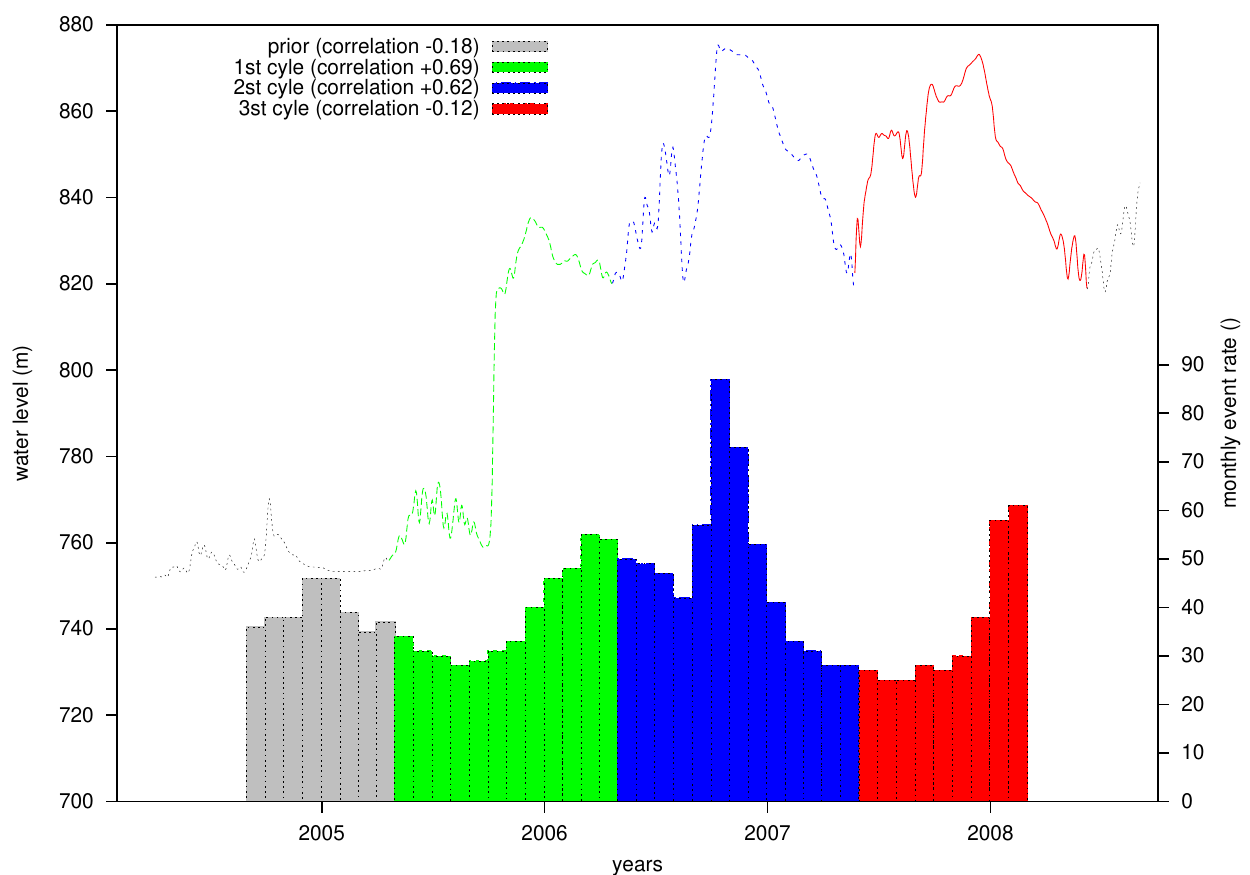}}
{\footnotesize {\bf Fig.~3.} Relationship between the water level change in the Minjiang river 
valley and the earthquake event rate, recorded in the vicinity of the water reservoir. The 
correlation coefficients are provided for each period prior and during the surface loading 
process.}
\end{minipage}
\end{center}

Second, from October 2005 until the mainshock nucleated in May 2008, $\Delta CFS$ exceeded triggering stresses 
of 4 kPa at $<$12 km depth. This stress level is critical, since it 
is daily generated by tidal elongations of the earth due to both the moon and the sun (see Appendix). 
It has been empirically shown that tidal stress changes have weak or random effects on triggering 
medium- to large-size earthquakes \cite{MaukKienle1973,Klein1976,BeelerLockner2003a,Cochran-etal2004}.
Thus, it can be anticipated that any triggering stress perturbation in the earth's crust must exceed 
at least stress levels of 1-10 KPa resulting from the tidal elongation of the earth. 

\begin{center}
\begin{minipage}[h]{10cm}
  \framebox{\includegraphics[width=10cm]{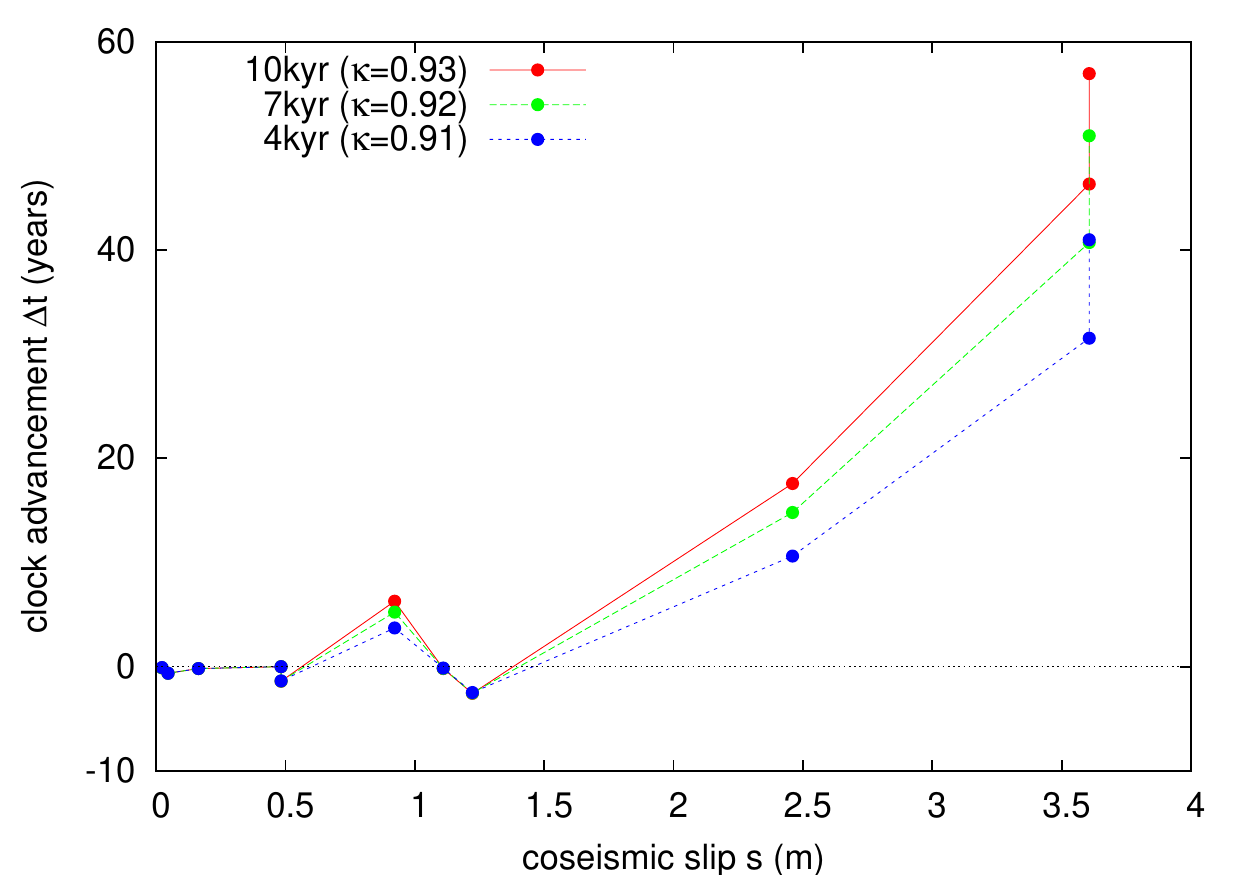}}
{\footnotesize {\bf Fig.~4.} Relationship between the coseismic slip $s$ observed during 
the mainshock \cite{Wang-etal2008a} and the clock advancement $\Delta t$ of static 
triggering stress perturbations as result of the surface loading 
processes of the artificial water reservoir (see Fig.~2). $s$ and $\Delta t$ are 
sampled from non-interpolated finite elements of the double-listric Beichuan fault 
beneath the reservoir. $\Delta t$-values are based on a rate-and-state dependent 
friction law (see supplementary material). Positive/Negative $\Delta t$-values indicate a 
fault de/stabilization. Correlation coefficients between $s$ and $\Delta t$ are indicated 
by $\kappa$.}
\end{minipage}
\end{center}

Third, the Zipingpu reservoir only provided the trigger for the limited, initial rupture on the thrust beneath 
the reservoir (Fig.~1c), which grew to engage the natural accumulated stress $\Delta\sigma$ along the full Beichuan 
system (BT1-3) of about 250 km length (Fig.~1b). $\Delta\sigma$ had been built up by natural tectonic 
loading during the seismic cycle along BT1-3 and the entire Longmen Shan. Moreover, the water reservoir also 
influenced the initial rupture propagation. In the first 20 seconds, the rupture propagation 
on a 32 km long and 13$\pm$3 km wide BT1-fault segment was directed upwards toward the Minjiang River Valley 
directly to the surface loading area of the Zipingpu water reservoir (Fig.~1b). Then, the rupture propagation 
process changed to a right-lateral NNE strike-slip rupture mechanism \cite{Zhang-etal2009,Wang-etal2008a}.
Thus, although expected and observed fault geometries differ, 
the expected seismic magnitude still falls, with an underestimation, into the statistical 
uncertainty of the observed magnitude.
Last, modeling results based on a rate-and-state dependent friction law (see Section \ref{sec:Gomberg})
suggest that the clock advancement $\Delta\,t$ 
of the mainshock varies for different earthquake cycles and model regimes (Tab.~\ref{model-results}). 
As discussed by Gomberg et al.~2000, the smaller $a$ is relatively to $b$, the more the 
fault becomes Coulomb-like. This might be true at 11-19 km depth, where faults are generally 
locked in the Longmen Shan region \cite{Wang-etal2008b}. Thus, regime 3 in 
Table \ref{model-results} might be more suitable to model the seismic cycle of the Beichuan 
faults, due to low deformation rates and the stiff behaving Pengguan Massif (Fig.~1).
Static perturbations of the water reservoir more likely produced $\leq$41 years of equivalent 
tectonic loading in a 4 kyr earthquake cycle or $\leq$57 years in a 10 kyr cycle. The more (less) the 
surface loading of the water reservoir perturbed BT1 and BT2 in its near vicinity 
(15 km radius) the more (less) it contributed to the coseismic slip observed on both 
faults during the main rupture (Fig.~4). This contribution, however, was $<$1$\%$. Although the 
static triggering seems to contribute only up to 6 decades of equivalent tectonic loading, it 
strongly correlates with the coseismic slip 
observed on BT1 and BT2 beneath the reservoir (Fig.~4). It should be noted that seasonal water level 
changes might advance the clock even further. According to the rate-and-state 
dependent friction law by Gomberg et al. (1998), dynamic stress perturbations applied late in the 
earthquake cycle, would advance the clock by several hundred years. Problematic, however, is that the 
friction law approximates the dynamic load as a box function instead of a wave function, which results in a
non-optimal solution. Thus, dynamic loads are not considered in this study. 

\begin{table}[!h]
	\centering
	\caption{Expected clock advancement times $\Delta\,t$ (in years) for the 2008 Wenchuan 
		due to static triggering stress perturbations $\Delta\,CFS$ of 
		about 2 kPa at hypocentral depth and 9 kPa at 4 km beneath the 
		Zipingpu water reservoir. Results are based on the rate-and-state 
		model suggested by Gomberg et al.~(1998) with respect to 
		a) three earthquake cycles and b) three model regimes describing the potential 
		behavior of the Beichuan fault.}
\begin{tabular}{ccccc}
\hline\hline
$t_f$	&$\Delta\,CFS$	&\multicolumn{3}{c}{static part}\\
years	&kPa 	&regime 1&regime 2&regime 3\\\hline\hline
4000	&9	&3	&13	&41\\
4000	&2	&1	&2	&4\\
7000	&9	&3	&14	&51\\
7000	&2	&1	&3	&5\\
10000	&9	&3	&15	&57\\
10000	&2	&1	&3	&6\\\hline\hline
	\end{tabular}
	\label{model-results}
\end{table}

\section{Conclusion}

This study suggests that surface loading of at least 320 million metric tons of 
water, which accumulated in the Zipingpu water reservoir in the Minjiang River Valley 
between 2005 and 2008, most likely triggered advanced China's Wenchuan M7.9 earthquake 
of May 12, 2008, while enhancing a reverse-fault rupture propagation within the first 20 seconds.
Specifically, 3-dimensional geomechanical modeling results based on a rate-and-state 
dependent friction law show that static triggering stresses brought parts of the Benchuan 
thrust fault system nearby the water reservoir closer to failure and advanced the clock 
of the mainshock by up to six decades. Conversely, other fault segments, directly beneath 
the reservoir, were brought away from failure due to the weight of the reservoir. 

The estimated slip of equivalent tectonic loading that was produced by the water reservoir 
on the Beichuan fault system indicates a strong correlation with the observed coseismic slip 
during the mainshock of May 12, 2008. The highest correlation coefficient of 0.93 was 
found for an earthquake cycle of 10 kyr. Moreover, correlations become weaker with decreasing 
recurrence time of the earthquake cycle. This confirms results of previous paleoseismicity 
studies, which show evidence that the Longmen Shan and, in particular, the Pengguan Massif 
is characterized by a $>$7 kyr seismic cycle for major M$>$7 earthquakes.

\section{Acknowledgments}
The author is grateful to Think Geohazards for its generous financial support. He also thanks the five anonymous reviewers for their 
constructive critiques and C.H.\ Scholz and L.\ Seeber from Lamont-Doherty Earth Observatory for their suggestions and comments to improve 
this manuscript.

\section{Appendix}

\label{sec:tides}
Everyday, the moon and sun cause tidal elongations on earth \cite{Bartels1957}.
These daily pull/push effects, in turn, induce stabilizing and destabilizing stresses on 
preexisting fault zones in the earth's crust and are independent from any geological forces on 
earth, including endogenous forces (e.g., volcanism, tectonics) and exogenous forces 
(e.g., erosion, sedimentation). Moreover, it has been shown that tidal stress changes have 
weak effects on triggering medium- to large-size earthquakes \cite{BeelerLockner2003a}. 
Thus, it can be anticipated that any triggering stress perturbation on the earth's crust must 
exceed at least stress levels resulting from the tidal elongation of the earth. 
Analytically, it can be shown how high these tidal stress changes are.

Let's assume, $F_0$ is the earth's gravitational potential, which results from both attraction 
force and centrifugal force of the earth and moon/sun. Tidal forces $V$, however, deform $F_0$:

\begin{equation}
F = F_0 - V
\end{equation}

This results in a vertical surface displacement $\xi$ with respect to the average value 
of the gravitation acceleration on earth $g$ = 9.798 ms$^{-1}$.

\begin{equation}
\xi = \frac{F_0 - F}{g} = \frac{V}{g}.
\end{equation}

Tidal forces change with geocentric zenith distance $\theta$ from the moon/sun \cite{Bartels1957}, 
whereas the main term of the tidal potential is 

\begin{equation}
V \approx \biggl( \frac{G}{\overline r_E^2} \biggr) r_E^2 \biggl( \cos 2\theta + \frac{1}{3} \biggr)
\end{equation}

with the lunar tidal constant $G_l$ = 2.6206 m$^2$s$^{-2}$ and the solar tidal constant 
$G_s$ = 1.2068 m$^2$s$^{-2}$, 
the radius of the earth $r_E$ and 
the mean radius of the earth $\overline r_E$ = 6371.221 km. Assuming the earth is a 
sphere ($r_E$ = $\overline r_E$), the general form of vertical surface displacement is  

\begin{equation}
\xi = \frac{G}{g} \biggl( \cos 2\theta + \frac{1}{3} \biggr).
\end{equation}

The displacement due to the moon and sun is

\begin{equation}
\xi_l = 0.267 {\text m} \biggl( \cos 2\theta + \frac{1}{3} \biggr), \;\;\; \xi_s = 0.123 {\text m} \biggl( \cos 2\theta + \frac{1}{3} \biggr).
\end{equation}

Thus, the $\xi_l$ and $\xi_s$ are amplified at the zenith ($\theta$ = 0$^\circ$) by 0.356 m and 0.164 m. 
On the other hand, they are depressed at the nadir ($\theta$ = 90$^\circ$) by 0.178 m and 0.082 m. 
The peak-trough difference for the moon is 0.534 m and 0.246 m for the sun.

Both vertical displacements induce maximal shear stresses $\tau$ and normal stresses $\sigma_n$ on 
preexisting faults (e.g., dipping 45$^\circ$) in the earth's crust with an average shear modulus 
of about $G$ = 30 GPa and friction angle of, let's assume, $\phi$ of 29$^\circ$:

\begin{equation}
\tau = \frac{\xi}{\overline r_E}\,2G,
\label{eq:HookSupp}
\end{equation}
\begin{equation}
\sigma_n = \frac{\tau}{\tan\phi}, 
\label{eq:CoulombSupp}
\end{equation}

Thus, maximum induced stresses that need to be exceeded by any additional triggering stress 
(e.g., due to surface loading) are:

\begin{center}
by the moon $\tau_l$ = 5.03 kPa and $\sigma_{n,l}$ = 9.07 kPa,
\end{center}
\begin{center}
by the sun $\tau_s$ = 2.32 kPa and $\sigma_{n,s}$ = 4.18 kPa.
\end{center}

\newpage \clearpage


\bibliographystyle{spbasic}      


\newpage

\newpage

\end{document}